\documentclass[amssymb,aps,prb,floatfix,floats,preprintnumbers,twocolumn,superscriptaddress,longnofootinbib,longbibliography]{revtex4-2}
\usepackage{rcs}
\usepackage{color,graphicx}
\usepackage{dcolumn}
\usepackage{bm}
\usepackage[normalem]{ulem}
\usepackage{chngcntr}
\usepackage[thinspace,thinqspace]{SIunits}
\usepackage{amsmath}
\usepackage{natbib}
\usepackage{hyperref}
\usepackage{xcolor}
\begin{document}
\preprint{APS/123-QED}
\title{Tetragonal-to-orthorhombic distortion in CeRh$_{2}$As$_{2}$}

\author{P. Khanenko}
\email[Corresponding author:~]{p.khanenko@hzdr.de}
\affiliation{Hochfeld-Magnetlabor Dresden (HLD-EMFL) and Würzburg-Dresden Cluster of Excellence ctd.qmat, Helmholtz-Zentrum Dresden-Rossendorf, 01328 Dresden, Germany}
\affiliation{Max Planck Institute for Chemical Physics of Solids, 01187 Dresden, Germany}
\author{M. Pfeiffer}
\affiliation{Max Planck Institute for Chemical Physics of Solids, 01187 Dresden, Germany}
\affiliation{Institute for Solid State and Materials Physics, Technical University Dresden, 01062 Dresden, Germany}
\author{K. Semeniuk}
\affiliation{Max Planck Institute for Chemical Physics of Solids, 01187 Dresden, Germany}
\affiliation{Institute for Quantum Materials and Technologies, Karlsruhe Institute of Technology, 76131 Karlsruhe, Germany}
\author{E. Hassinger}
\affiliation{Max Planck Institute for Chemical Physics of Solids, 01187 Dresden, Germany}
\affiliation{Institute for Quantum Materials and Technologies, Karlsruhe Institute of Technology, 76131 Karlsruhe, Germany}
\author{P. Thalmeier}
\affiliation{Max Planck Institute for Chemical Physics of Solids, 01187 Dresden, Germany}
\author{J. Wosnitza}
\affiliation{Hochfeld-Magnetlabor Dresden (HLD-EMFL) and Würzburg-Dresden Cluster of Excellence ctd.qmat, Helmholtz-Zentrum Dresden-Rossendorf, 01328 Dresden, Germany}
\affiliation{Institute for Solid State and Materials Physics, Technical University Dresden, 01062 Dresden, Germany}
\author{A. M. Le\'on}
\affiliation{Departamento de Física, Facultad de Ciencias, Universidad de Chile, Casilla 653, Santiago, Chile}
\author{G. Zwicknagl}
\affiliation{Institute for Mathematical Physics, Technische Universit\"at Braunschweig, D-38106 Braunschweig, Germany}
\affiliation{Max Planck Institute for Chemical Physics of Solids, 01187 Dresden, Germany}
\author{C. Geibel}
\affiliation{Max Planck Institute for Chemical Physics of Solids, 01187 Dresden, Germany}
\author{S. Khim}
\affiliation{Max Planck Institute for Chemical Physics of Solids, 01187 Dresden, Germany}
\author{M. Brando}
\email[Corresponding author:~]{brando@cpfs.mpg.de}
\affiliation{Max Planck Institute for Chemical Physics of Solids, 01187 Dresden, Germany}
\date{\today}
\begin{abstract}
CeRh$_{2}$As$_{2}$ is best known for its two superconducting phases, which are observed when a magnetic field is applied along the $c$ axis of its tetragonal unit cell. The emergence of superconductivity, with $T_{c} \approx 0.31$\,K, is preceded by an ordered state (phase I), with an ordering temperature $T_{0} \approx 0.46$\,K. The precise nature of the order parameter in phase I is still unknown, although evidence of dipolar ordering was found. To investigate a possible symmetry breaking at $T_{\textrm{0}}$, we conducted longitudinal thermal-expansion measurements along the crystallographic [100], [010], [110], and [1$\overline{1}$0] directions on a single-crystalline sample cut into an octagonal shape. Our results demonstrate the breaking of the fourfold in-plane symmetry below $T_{\textrm{0}}$, evidencing a small tetragonal-to-orthorhombic distortion (orthorhombicity $\delta \sim 10^{-7}$).
The breaking of the tetragonal symmetry at $T_{\textrm{0}}$ might imply that the superconducting order parameter that develops within phase I cannot be multi-component.
More generally, the distortion places restrictions on the possible order parameters for phase I.
\end{abstract}

\maketitle

The heavy-fermion (HF) superconductor CeRh$_{2}$As$_{2}$ has attracted significant interest in recent years due to the presence of two superconducting (SC) phases appearing in a magnetic field applied along the $c$ axis of its tetragonal unit cell~\cite{khim2021,landaeta2022a}. This discovery makes CeRh$_{2}$As$_{2}$ one of the few established multiphase superconductors alongside UPt$_{3}$~\cite{bruls1990, adenwalla1990}, thorium-doped UBe$_{13}$~\cite{ott1985}, UTe$_{2}$~\cite{ran2019, braithwaite2019}, and YbRh$_{2}$Si$_{2}$~\cite{levitin2026}. In contrast to spin-triplet superconductors like UPt$_{3}$, the multiple SC phases in CeRh$_{2}$As$_{2}$ are possibly caused by its locally non-centrosymmetric crystalline structure (space group P4/nmm), which allows for strong Rashba spin-orbit coupling at the Ce atoms, staggered along the $c$ axis. Consequently, applying a magnetic field along the $c$ axis can induce a transition from an even-parity to an odd-parity SC state, in which the SC order parameter (OP) changes sign between the layers along the $c$ axis~\cite{khim2021,yoshida2012,sigrist2014,moeckli2021,cavanagh2022,nogaki2022}. An in-plane field would leave the SC state Pauli limited, as observed~\cite{khim2021}.

An alternative, more recent scenario considers the possibility of a transition between coexistence phases of the same SC OP with distinct admixtures~\cite{jakubczyk2026}. Other theoretical models rely on ARPES measurements~\cite{chen2024a,chen2024b,wu2024} and on van Hove singularities in the band structure of the nonsymmorphic crystal structure~\cite{lee2025,nogaki2026}, in particular at a wave vector $(\pi,\pi)$, along which AFM spin correlations were observed in neutron scattering~\cite{chen2024}. However, ARPES measurements were only carried out at high temperatures, i.e., near but not well below the Kondo temperature of about 30\,K, where the heavy bands form. ARPES experiments are also challenging in Ce-based HF compounds due to the small resolution and the fact that they are limited only to the occupied states of the Kondo resonance~\cite{denlinger2001,koitzsch2013}.

Importantly, in all parity-transition models based on Rashba spin-orbit coupling, the SC OP can have a single component with spin-singlet coupling in both SC states. It acquires different parity due to the lattice degrees of freedom. While the precise nature of the SC OP remains unknown, NMR experiments have already indicated the presence of spin-singlet superconductivity in both SC phases~\cite{kibune2022,ogata2023,ogata2024}, and the analysis of ultrasound experiments suggests that the SC OP is single component in both SC phases~\cite{galeski2026}.

The situation in CeRh$_{2}$As$_{2}$ is further complicated by the presence of an ordered state (phase I), within which superconductivity emerges. At low temperatures, CeRh$_{2}$As$_{2}$ is in a HF non-Fermi liquid (NFL) state and the transition into phase I occurs at $T_{\textrm{0}}$ $\approx 0.46$\,K~\cite{khim2021,hafner2022,khanenko2025a,khanenko2025b}, while superconductivity sets in at $T_{\textrm{c}}$$\ \approx 0.31$\,K. When a magnetic field is applied along the $c$ axis, $T_{\textrm{0}}$ is continuously suppressed and reaches zero at about 7\,T, whereas the even-parity SC1 phase changes into the odd-parity SC2 phase at a critical field of $H^{\mathrm{*}} \approx 4$\,T. The SC2 state is orbitally limited with a very large upper critical field of approximately 14\,T~\cite{chajewski2024,semeniuk2023,khanenko2025b}. In contrast, when the field is applied within the $ab$ plane, only the Pauli-limited SC1 state is present and $T_{\textrm{c}}$ is suppressed at around $2$\,T. Conversely, for this field direction, $T_{\textrm{0}}$ changes only marginally in fields up to 3\,T, after which it increases with increasing field. A strong first-order phase transition into another ordered state (phase II) occurs at a critical field $\mu_{0}$$H_{\textrm{cr}}$ which is between 6 and 9\,T, depending on the precise in-plane direction along which the magnetic field is applied~\cite{semeniuk2026}. The transition temperature $T_{\textrm{1}}$ into phase II increases rapidly with increasing field, indicating a predominant quadrupolar origin of the order parameter in this phase~\cite{mishra2022,schmidt2017,thalmeier2025,schmidt2024,khanenko2025a,numa2026}. At even higher magnetic fields, additional features in the magnetoresistance have been observed~\cite{blawat2025}.

The nature of the phases I and II is still unclear. Recent $\mu$SR~\cite{khim2025}, NMR~\cite{kibune2022,ogata2024,ogata2026}, and micro-Hall magnetometry~\cite{juraszek2025} experiments on high-quality single crystals have detected internal static magnetic fields setting in below $T_{\textrm{0}}$, pointing to dipolar ordering in phase I. However, the phase diagram for fields in the $ab$ plane is not compatible with phase I being simply an antiferromagnetic order, but requires higher-order-multipole degrees of freedom. Two scenarios have been proposed: i) An ordering of the local dipolar OP in zero magnetic field, which develops a concomitant quadrupolar OP at higher fields~\cite{schmidt2024,thalmeier2025} and ii) a fully itinerant ordering due to nesting of the heavy bands, which results in a predominant quadrupolar density wave state~\cite{hafner2022}. None of the scenarios, however, can account for all observations reported for phases I and II.

The presence of phase I prompted new proposals as to understand the multiphase superconductivity, such as a metamagnetic transition taking place at $H^{*}$~\cite{machida2022}. This was, however, ruled out by an analysis of the phase diagram~\cite{semeniuk2023,khanenko2025b} and pressure experiments~\cite{pfeiffer2024,semeniuk2024}, which indicate that phase I interacts only weakly with superconductivity and is irrelevant to the parity transition. 

\begin{figure}[t]
	\begin{center}
		\includegraphics[width=\columnwidth]{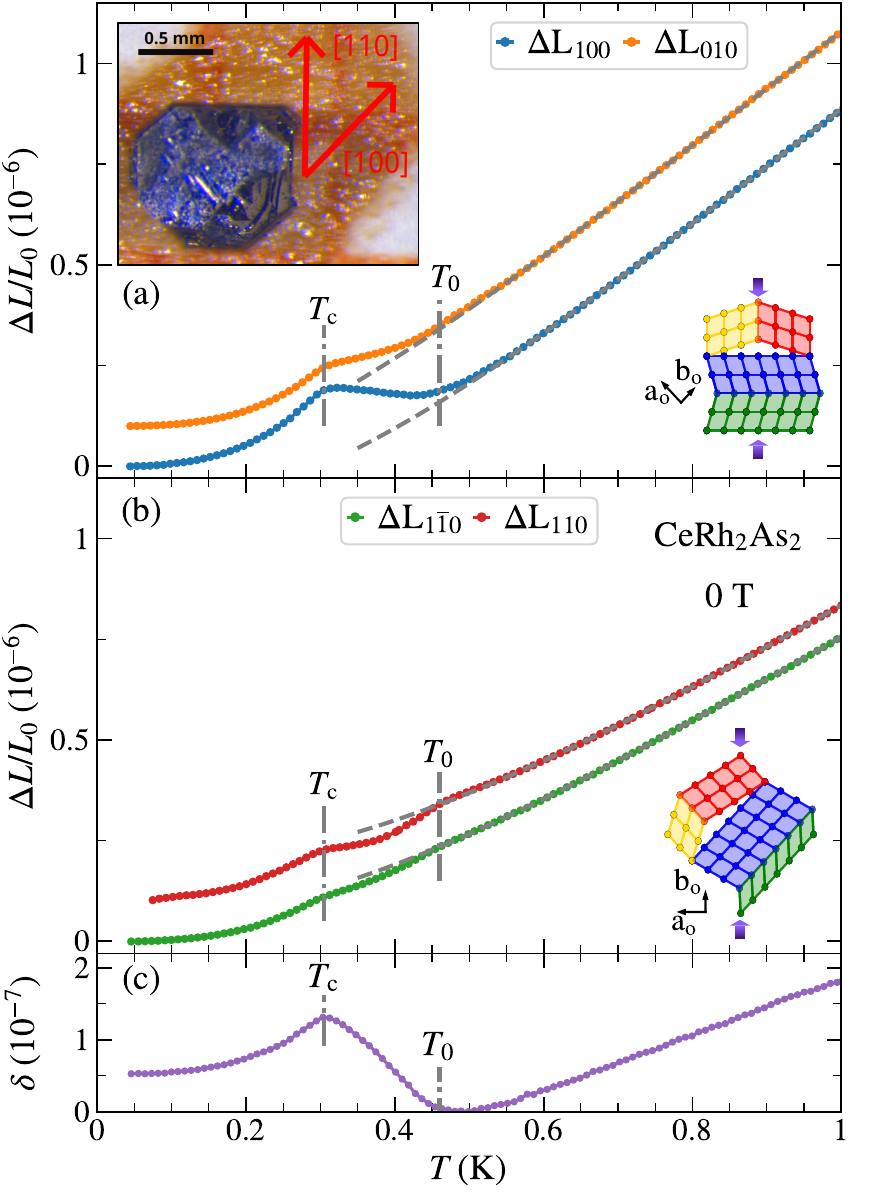}
		\caption{Zero-field uniaxial thermal expansion $\Delta L(T) / L_{0}$ of CeRh$_{2}$As$_{2}$ for the crystallographic directions (a) [100], [010] and (b) [110], [1$\overline{1}$0]. For clarity, the curves are shifted by $0.1\times 10^{-6}$ with respect to each other. See the main text for details on the dashed and dash-dotted lines. (a) The inset shows the sample used in this study, which was cut into an octagonal shape. In the lower right corners there are illustrations about the potential influence of uniaxial pressure applied along (a) [100], [010] and (b) [110], [1$\overline{1}$0] on the orthorhombic domains of the $B_{2g}$ orthorhombic distortion. Different colors indicate different domains, $a_{o}$ and $b_{o}$ are shown for the blue domain. (c) Orthorhombicity $\delta$ calculated using $\Delta \mathrm{L}_{100}$ and $\Delta \mathrm{L}_{1\overline{1}0}$ shown in panels (a) and (b), see text for detail.}
		\label{fig1}
	\end{center}
\end{figure}

In ultrasound measurements, in contrast, phase I does influence the signatures at $T_{\textrm{c}}$ in the elastic constant $C_{66}$. At ambient pressure, a step-like transition is observed at $T_{\textrm{c}}$, suggesting a possible two-component SC OP. However, when phase I is suppressed by pressure, only a kink is observed at $T_{\textrm{c}}$, proving that the SC OP is single component~\cite{galeski2026}. This behavior could be simply related to a distortion of the lattice at $T_{\textrm{0}}$, e.g., from tetragonal to orthorhombic, also supported by a very strong drop of $C_{66}$ at $T_{\textrm{0}}$, three times larger than the signature in the respective constant $(C_{11}-C_{12})/2$~\cite{galeski2026}. In fact, the elastic constant $C_{66}$ corresponds to the shear $\epsilon_{xy}$, which couples to the local susceptibility of the $O_{xy}$ quadrupole operator~\cite{thalmeier1991}. It is worth mentioning that, since $C_{66}$ shows the strongest anomaly at $T_0$, we can assume that $\epsilon_{xy}$ itself primarily couples to the magnetic order parameter $\bf M$. Due to the time-reversal invariance, in the free energy expression, the strain can only be multiplied by an even number of $M_\alpha$ components $(\alpha=x,y,z)$, and at the same time, the product of the components must have the same symmetry as $\epsilon_{xy}$. Hence, the leading coupling term in the free energy must be $\sim\epsilon_{xy}M_xM_y$, which requires both components $M_x$ and $M_y$ to be nonzero. Since the directions $x$ and $y$ are equivalent above $T_0$, this indicates that the magnetic moments should have a component along the $[110]$ direction.

Another important indication of a possible lattice distortion is based on the analysis of the Ehrenfest relation $dT_{0}/dp = V_{m}T_{0}(\Delta \beta/\Delta C)$ reported in Ref.~\cite{hafner2022}, where $V_{m}$ is the molar volume, $\Delta \beta$ is the jump in the volume thermal-expansion coefficient and $\Delta C$ is the jump in the heat capacity at $T_{\textrm{0}}$. A positive $dT_{0}/dp$ was estimated, while the pressure experiments showed the opposite~\cite{pfeiffer2024}. Since $\Delta\beta(T)$ was calculated by using the thermal-expansion coefficients only along [001] and [110], $\Delta\beta(T) = \Delta\alpha(T)_{[001]} + 2\Delta\alpha(T)_{[110]}$, in case of a lattice distortion, e.g., of $B_{2g}$ type, this equation alone would not allow a proper determination of $\Delta\beta$ at the transition.

In order to investigate a possible symmetry breaking at $T_{\textrm{0}}$, we conducted longitudinal thermal-expansion measurements along the crystallographic [100], [010], [110], and [1$\overline{1}$0] directions on a single-crystalline sample cut into an octagonal shape (cf. Fig.~\ref{fig1}). Our results demonstrate the breaking of the fourfold in-plane symmetry below $T_{\textrm{0}}$, indicating a small (orthorhombicity $\delta \sim 10^{-7}$) tetragonal-to-orthorhombic distortion of the $B_{2g}$ type, with the formation of orthorhombic domains. The orthorhombic phase coincides with phase I. The breaking of the tetragonal symmetry at $T_{\textrm{0}}$, if large enough, implies that the superconducting order parameter that develops at SC1 cannot be multi-component because of the underlying orthorhombic structure. 
%

The CeRh$_{2}$As$_{2}$ single crystal used for our experiments was grown using a Bi-flux method~\cite{khim2021}. Thermal-expansion and magnetostriction measurements were carried out in a dilution refrigerator, using a commercially available capacitance dilatometer~\cite{kuechler2012} at temperatures down to 50\,mK and in magnetic fields up to 10\,T. The SC and phase I transition temperatures indicate that the crystal is of intermediate quality and belongs to group 2 according to the classification of Ref.~\cite{khanenko2025b}. The sample was cut into the  octagonal shape using a wire saw to enable measurements along the primary axes and diagonals within the $ab$ plane. The sample is shown in the inset of Fig.~\ref{fig1}(a).  Using Laue diffraction, we determined that the cuts deviate from the crystallographic directions by not more than $7$ ($10$ for the [1$\overline{1}$0] direction) degrees.
\begin{figure}[t]
	\begin{center}
		\includegraphics[width=\columnwidth]{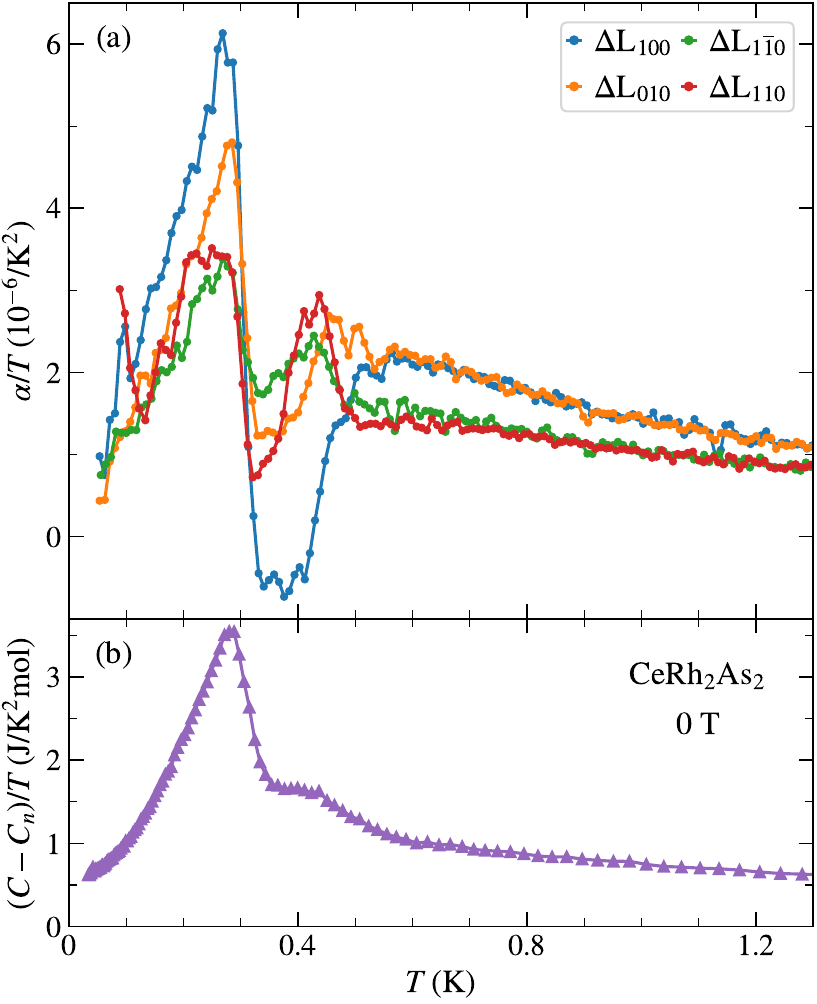}
		\caption{(a) Zero-field thermal-expansion coefficient divided by temperature $\alpha(T)/T$ of the four measured crystallographic directions, calculated using the data shown in Figs.~\ref{fig1}(a) and ~\ref{fig1}(b). (b) Electronic specific heat $(C - C_{n})/T$ of CeRh$_{2}$As$_{2}$, with subtracted nuclear heat capacity (see Ref.~\cite{semeniuk2023} for details) of a sample of similar quality to that used in this study.}
		\label{fig2}
	\end{center}
\end{figure}

Figure~\ref{fig1} summarizes the most important result of our work: It shows the thermal expansion $\Delta L /L_{0}$ measured along the equivalent directions [100] and [010] (Fig.~\ref{fig1}(a)) and along the equivalent directions [110], and [1$\overline{1}$0] (Fig.~\ref{fig1}(b)) in zero field. The dashed lines are extrapolations of polynomial fits to the high-temperature data (0.6 to 1.4\,K) down to 0.35\,K. $T_{\textrm{c}}$\ $ \approx 0.31$\,K and $T_{\textrm{0}}$ $ \approx 0.46$\,K are indicated by the vertical dash-dotted lines. They are extracted as the middle point at the jump in the thermal-expansion coefficient $\alpha = 1/L_{0}\ \partial(\Delta L)/\partial T$ (shown in Fig.~\ref{fig2}(a)), and averaged over all measurements shown in Figs.~\ref{fig1}(a) and (b). At $T_{\textrm{c}}$, we observe a clear downturn in $\Delta L /L_{0}$ for all four directions. At $T_{\textrm{0}}$, on the contrary, there is a clear downward kink in $\Delta L_{[110]}$ and $\Delta L_{[1\overline{1}0]}$ but an upturn in $\Delta L_{[100]}$ and $\Delta L_{[010]}$. This is even better visible in the plot of $\alpha/T$ vs $T$ (Fig.~\ref{fig2}(a)). In a tetragonal system, such as CeRh$_{2}$As$_{2}$, one expects an isotropic response of the thermal expansion within the $ab$ plane. The deviation from this expected behavior unambiguously shows that the transition to phase I is accompanied by a lattice distortion.

The behavior observed at $T_{\textrm{0}}$ immediately explains why in Ref~\cite{hafner2022} the estimation of $dT_{0}/dp$ from the Ehrenfest relation was incorrect. In that work, only thermal expansion along [110] was considered, and the in-plane thermal expansion was assumed to be isotropic, which is not true for an orthorhombic distortion. The expansion along [100] shows that the crystal expands within the $ab$ plane. In order to calculate the volume change at $T_{\textrm{0}}$, and so the correct Ehrenfest relation, measurements on a single-domain sample would be needed. However, in our measurements the sample was not in a single domain below $T_{\textrm{0}}$. In fact, the magnitude of the expansion below $T_{\textrm{0}}$ in the [100] direction is different from the one along the [010], although these are equivalent directions. The same is observed for [110] and [1$\overline{1}$0]. This demonstrates that the sample is not single domain within phase I and, therefore, it is impossible to calculate $dT_{0}/dp$ precisely from our measurements alone. Moreover, in the capacitive dilatometer cell used in this study, a force of approximately 3\,N~\cite{kuechler2012} is applied to the sample, which has a cross-sectional area of about $\sim 1$\,mm$^{2}$. The resulting uniaxial pressure of 3\,MPa can change the domain distribution within the sample during every measurement. For instance, it has been shown that 0.8\,MPa can reverse the sign of $\alpha(T)$ at the structural transition in BaCo$_{2}$V$_{2}$O$_{8}$~\cite{niesen2013}, but much higher pressure is needed to fully detwin the crystal. In iron pnictides, however, at least 5-10\,MPa are needed to detwin a crystal~\cite{fisher2011}. In our case the sample remained twinned during the measurements.
\begin{figure}[t]
	\begin{center}
		\includegraphics[width=\columnwidth]{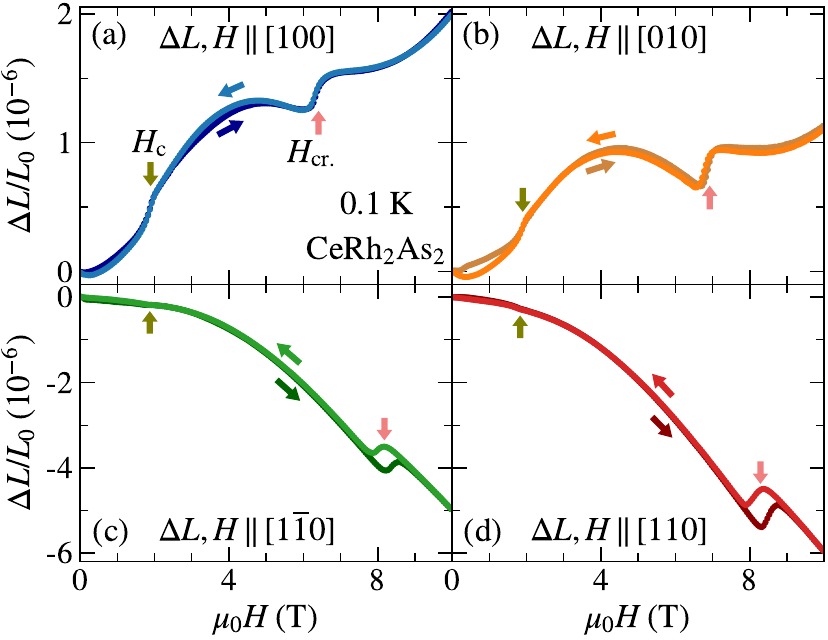}
		\caption{Longitudinal uniaxial magnetostriction $\Delta L(H) / L$ of CeRh$_{2}$As$_{2}$ at 100\,mK along the directions (a) [100], (b) [010], (c) [1$\overline{1}$0], and (d) [110]. Both, up- and down-sweeps are shown. The sweep directions are indicated by arrows of the corresponding color. The vertical olive and pink arrows indicate the SC-to-phase I and phase I-to-phase II transitions at $H_{\textrm{c}}$ and $H_{\textrm{cr}}$, respectively.}
		\label{fig3}
	\end{center}
\end{figure}

We now discuss the presence of orthorhombic domains in phase I. For a $B_{1g}$-type transition at $T_{\textrm{0}}$, in the absence of domains, we would expect the sample to shrink (expand), relative to the background, along the [100] direction and expand (shrink) along the [010] direction. This, however, does not correspond to our results. For a $B_{2g}$-type transition, we would expect the same behavior for the [110] and $[1\overline{1}0]$ directions, respectively, which again is not what we observe. A similar situation has been reported in some iron pnictides, in which a tetragonal-to-orthorhombic transition occurs with the concomitant formation of orthorhombic domains (Ref.~\cite{fisher2011} and references therein). More precisely, there are four types of domains possible in the orthorhombic phase. These domains differ in the orientation of the primitive in-plane orthorhombic vectors $a_{o}$ and $b_{o}$, $a_{o} > b_{o}$ and $a_{o} \perp b_{o}$. In the absence of external uniaxial pressure, the domains should be distributed evenly throughout the sample. When in-plane uniaxial pressure is applied, along or perpendicular to the primitive vectors, the domains with $b_{o}$ parallel to the pressure are favored~\cite{fisher2011}, resulting in a shrinkage of the sample in longitudinal measurement. When pressure is applied at 45$^{\circ}$ to $a_{o}$ and $b_{o}$, the domain structure of the sample remains largely unaltered. This is schematically illustrated in the lower right corners of Figs.~\ref{fig1}(a) and ~\ref{fig1}(b) for a $B_{2g}$-type distortion. From the suppression of phase I under pressure and the associated Ehrenfest relation, we expect the volume of the sample to expand at $T_{\textrm{0}}$. The fact that in our measurements the sample shrinks along [110] and $[1\overline{1}0]$ indicates that the vectors $a_{o}$ and $b_{o}$ are oriented along these directions, and the uniaxial pressure induces domains with $b_{o}$ parallel to the applied pressure. Thus, the orthorhombic transition in CeRh$_{2}$As$_{2}$ is of the $B_{2g}$ type. This conclusion is supported by a large, in comparison to other modes, anomaly in the elastic mode $C_{66}$ at $T_{\textrm{0}}$, which corresponds to the $\epsilon_{B_{2g}}$ shear strain~\cite{galeski2026}. 

To estimate the size of the distortion, we calculate the orthorhombicity $\delta = (a - b) / (a + b)$. For that we must take into account the presence of the domains and the influence of the uniaxial pressure, as it was done, e.g., for Ba$_{1-x}$K$_{x}$Fe$_{2}$As$_{2}$~\cite{boehmer2015}. Since the transition is of the $B_{2g}$ type, $\Delta L_{[110]}$ and $\Delta L_{[1\overline{1}0]}$ correspond to the expansion of the $b$ axis due to the applied pressure, while $\Delta L_{[100]}$ and $\Delta L_{[010]}$ correspond to the expansion along the diagonal for an unaltered domain structure. Therefore, we can choose $\Delta L_{[1\overline{1}0]}$ as $b$-axis expansion and infer $a$ from $\Delta L_{[100]}$. The resulting $\delta$ is shown in Fig.~\ref{fig1}c. This calculation should be treated with caution caused by an underestimation of the absolute values due to incomplete detwinning of the sample during $\Delta L_{[1\overline{1}0]}$ measurements, small misalignments, and slightly differing uniaxial pressures in each measurement. Qualitatively, however, we can conclude that the tetragonal-to-orthorhombic transition is small, of the order of $\delta \sim 10^{-7}$ between $T_{\textrm{0}}$ and $T_{\textrm{c}}$.
\begin{figure}[t]
	\begin{center}
		\includegraphics[width=\columnwidth]{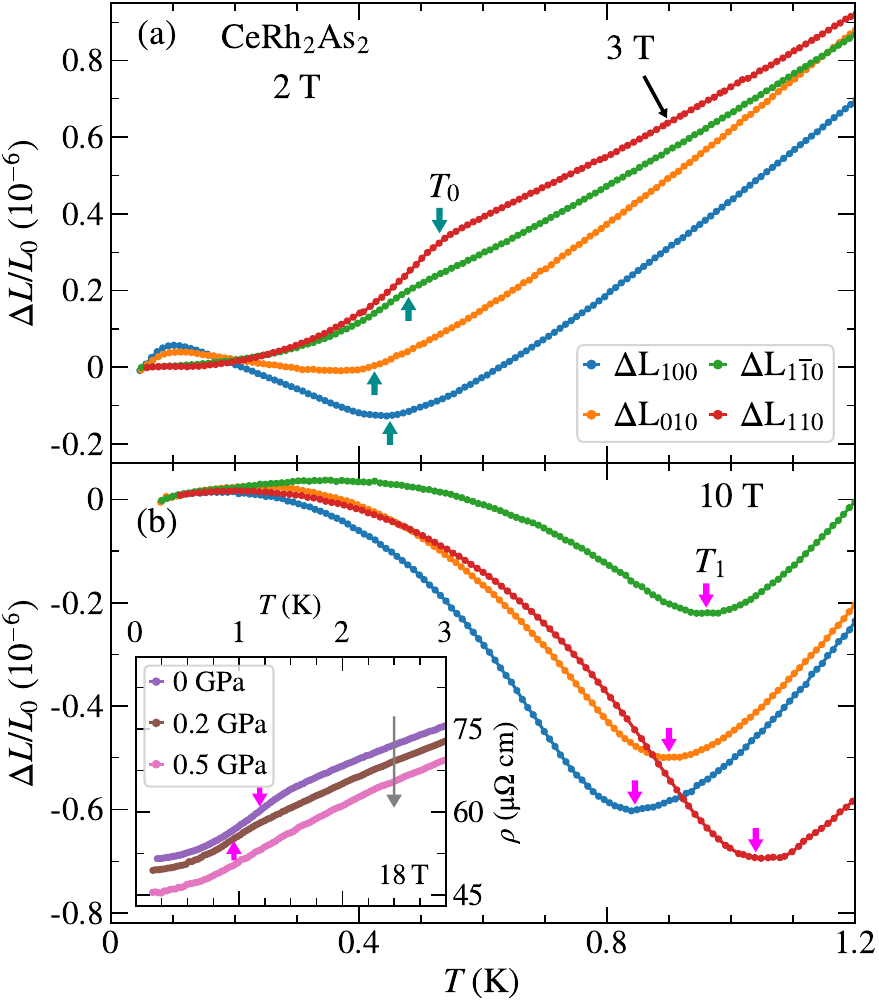}
		\caption{Longitudinal thermal expansion $\Delta L(T) / L_{0}$ of CeRh$_{2}$As$_{2}$ along [100], [010] [110], and [1$\overline{1}$0] at (a) 2\,T and (b) 10\,T. $\Delta L_{110} $ in (a) was measured at 3\,T. (a) The turquoise and (b) magenta arrows indicate the PM-to-phase I and PM-to-phase II transitions, respectively. The inset in (b) shows the temperature dependence of the electrical resistivity $\rho(T)$ at different hydrostatic pressures, with 18\,T applied in the $ab$ plane. The curves have been shifted by $-2\ \mathrm{ \mu \Omega \cdot cm} $ for clarity, as indicated by the gray arrow.}
		\label{fig4}
	\end{center}
\end{figure}

When calculating the thermal expansion coefficient $\alpha(T)$ from the data shown in Figs.~\ref{fig1}(a) and ~\ref{fig1}(b), we observe an intriguing behavior for $T > T_{0}$ (Fig.~\ref{fig2}(a)). For a simple metal or a HF compound, we expect a Fermi-liquid behavior with constant $\alpha(T)/T$ for $T \rightarrow 0$~\cite{zhu2003}. But, in all measurements, we observe an increasing $\alpha(T)/T$ with decreasing temperature very similar to the behavior found in the specific heat (Fig.~\ref{fig2}(b))~\cite{khim2021,hafner2022}. This NFL behavior has been associated to the presence of a nearby quantum critical point (QCP)~\cite{khim2021,pfeiffer2024,khanenko2025a}. The origin of this point is still under debate: It has been proposed that the QCP in CeRh$_{2}$As$_{2}$ is antiferromagnetic in nature, as in many other Ce-based HF systems, and, therefore, associated to the dipolar nature of phase I. This QCP is located under pressure at 0.5\,GPa. However, the SC dome and the NFL behavior extend over a much larger pressure range than in other Ce-based HF superconductors~\cite{pfeiffer2024}. Moreover, specific-heat measurements at high fields indicate that the NFL behavior is robust against field, so that it cannot be associated with only dipolar fluctuations, but possibly with multipolar fluctuations~\cite{khanenko2025a,numa2026}. 

In our measurements, we observe a different temperature dependence of $\alpha(T)/T$ for measurements along [100] and [010] compared to those along [110] and $[1\overline{1}0]$. Since the system is tetragonal above $T_{\textrm{0}}$ this behavior is very peculiar. It cannot result from isotropic fluctuations, like the two-dimensional magnetic fluctuations with easy-plane $XY$-type anisotropy observed in NMR experiments~\cite{kitagawa2022}. However, we associate this behavior with nematic fluctuations already present above the ordered state. In fact, a nematic character of the AFM spin correlations was observed in neutron-scattering experiments~\cite{chen2024}.

We now discuss the response of the system to magnetic field. To verify the orientations along which we cut the sample, we conducted longitudinal magnetostriction measurements at 100\,mK for all four directions, with field $H \parallel \Delta L$ up to 10\,T, and compared them with the study reported in Ref.~\cite{semeniuk2026}. The results are shown in Fig.~\ref{fig3}. We observe a kink in $\Delta L/L_{0}(H)$ vs $H$ at the SC critical field $H_{\textrm{c}}$ and a pronounced step with hysteresis at the critical field $H_{\textrm{cr}}$ of the transition from phase I to phase II. These critical fields, indicated by vertical arrows, were extracted from the magnetostriction coefficient $\lambda = 1/L_{0}\ \partial(\Delta L)/\partial H$ (averaged between up and down sweeps). By comparing the angle dependence of $H_{\textrm{cr}}$ determined in the in-plane anisotropy study of CeRh$_{2}$As$_{2}$~\cite{semeniuk2026} with our results, we can confirm that the misalignment of our cuts is below $7$ ($10$ for the [1$\overline{1}$0] direction) degrees. 

We now take a look at the temperature dependent $\Delta L$ at constant magnetic fields across $T_{\textrm{0}}$ (Fig.~\ref{fig4}(a)) and $T_{\textrm{1}}$ (Fig.~\ref{fig4}(b)). At small fields (2 to 3\,T), but strong enough to suppress the SC state, we observe the same behavior as in zero field with a tetragonal-to-orthorhombic distortion below $T_{\textrm{0}}$. So, we can state that the orthorhombic distortion is present in phase I also in magnetic field. On the other hand, at fields larger than $H_{\textrm{cr}}$, the sample expands in all directions below $T_{\textrm{1}}$. Importantly, although there is almost no difference between the measurements along $[100]$ and $[010]$, there is a large difference in magnitute between those along [1$\overline{1}$0] and $[110]$. For this reason we cannot exclude a possible $B_{2g}$-type distortion also when entering phase II. Using the Ehrenfest relation to estimate the pressure dependence of $T_{\textrm{1}}$ (see Fig.~S2 in SM), and the transverse thermal expansion along [001]~\cite{khanenko2025t}, we obtain $dT_{1}/dP = -0.5 \pm 0.1$\,K/GPa, which is almost half of the experimental value of $-1.25 \pm 0.2$\,K/GPa, extracted from resistivity measurements under pressure (see inset of Fig.~\ref{fig4}).

To investigate the interplay between structural, magnetic, and quadrupolar degrees of freedom in CeRh$_2$As$_2$, we performed density-functional-theory (DFT) and renormalized-band-structure (RBS) calculations (see SM for details). DFT identifies an interlayer antiferromagnetic ground  state with the Ce moments oriented along the [110] direction, preserving the tetragonal structure, whereas moments pointing along the [100] favor a $B_{1g}$ orthorhombic distortion (see Figs.~S3 and S4 in SM). This suggests that the experimentally observed $B_{2g}$ distortion cannot be driven by only dipolar magnetic degrees of freedom.
\begin{figure}[b]
	\begin{center}
		\includegraphics[width=\columnwidth]{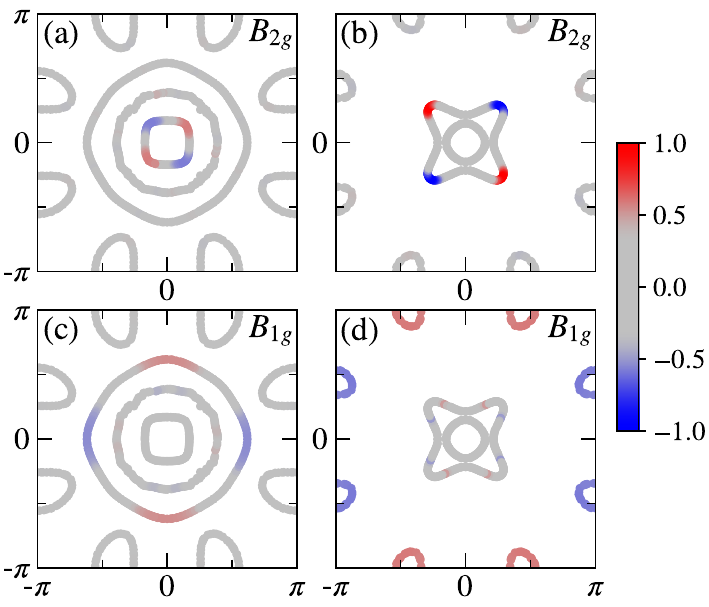}
		\caption{Sensitivity of the quasiparticle energies $\left|n\mathbf{k}\right>$ for states at the Fermi energy in the basal plane (a), (c) and top plane (b), (d) with respect to a uniform (“ferro”) perturbation with $B_{2g}$ (a), (b) and  $B_{1g}$ (c), (d) symmetries. The data are normalized to the maximum value on the large $\Gamma$-centered Fermi surface (the yellow ``onion'' in Fig.~5 of Ref.~\cite{hafner2022}).}
		\label{fig5}
	\end{center}
\end{figure}
To assess the role of quadrupolar degrees of freedom, we consider the same RBS calculations as used in Ref.~\cite{hafner2022}. Lattice distortions affect the heavy 4$f$-derived quasiparticles by two mechanisms~\cite{fulde1988,zwicknagl1992}: A local volume change leads to a change in the characteristic Kondo scale as described by the electronic Gr\"uneisen parameter. A change in the local symmetry at the Ce site, on the other hand, modifies the crystalline electric field (CEF). It is this effect we consider here and evaluate the response of the renormalized bands to homogeneous $B_{2g}$ and $B_{1g}$ perturbations of the CEF at the Ce sites. Figure~\ref{fig5} shows the resulting first-order energy shifts of the quasiparticle states at the Fermi level. The response to the $B_{2g}$ perturbation is strongly momentum dependent and changes sign along the Fermi surface, implying a deformation rather than a rigid energy shift. Remarkably, the Fermi sheet at the Brillouin-zone boundary (brown corrugated cylinders in Fig.~5 of Ref.~\cite{hafner2022}), where superconductivity has been proposed to reside, is less sensitive to the perturbation. The response is also strongly symmetry selective: the Fermi-level states at the large $\Gamma$-centered surface (the yellow ``onion'' in Fig.~5 of Ref.~\cite{hafner2022}) exhibit a much stronger overall sensitivity to the $B_{2g}$ perturbation than to the $B_{1g}$ perturbation.

To conclude, our longitudinal thermal-expansion measurements of a high-quality single crystal of CeRh$_{2}$As$_{2}$ along the crystallographic directions $[100]$, $[010]$, $[110]$, and [1$\overline{1}$0] unambiguously demonstrates the breaking of the four-fold in-plane symmetry of the tetragonal unit cell below $T_{\textrm{0}}$, evidencing that a tetragonal-to-orthorhombic distortion takes place at $T_{\textrm{0}}$. The analysis of the data suggests that the distortion is of $B_{2g}$ type, with the presence of orthorhombic domains. This distortion places restrictions on the possible order parameters for phase I. The change in the orthorhombicity $\delta$ is of the order of $10^{-7}$, which is a small value compared to orthorhombic transitions in other materials, such as in the iron pnictides/selenides (e.g., FeSe~\cite{boehmer2013}). It may be compared to the reported $\delta \sim 10^{-5}$ in URu$_{2}$Si$_{2}$~\cite{tonegawa2014, choi2018}, where the ordering temperature is two orders of magnitude larger. Now we can ask if this distortion is sufficiently large to split a multicomponent SC OP to lead to different $T_{\textrm{c}}$. This is maybe not the case, since this distortion is negligible when compared to a real orthorhombicity of chemical origin which is in the percent range. For example, in UPt$_3$ which undergoes an AFM transition at about 5\,K~\cite{aeppli1988}, this is not sufficient. The magnetic order reduces the sixfold symmetry to an orthorhombic one and the symmetry breaking is strong enough to ensure a single-component order parameter at the first SC transition at $T_{c1}$. Nevertheless, at a somewhat lower $T_{c2}$ the second component reappears leading to a multicomponent superconducting phase, which can approximately be classified as a degenerate $E$ representation of the high-symmetry group D$_{6h}$. Therefore, it is very much possible that also in CeRh$_{2}$As$_{2}$ a small lattice distortion does not fundamentally change the classification of the SC order parameter according to high-symmetry group and that the SC OP of phase SC1 is single component, in agreement with the conclusion reported in Ref.~\cite{galeski2026}. Since it has been shown that phase I is suppressed under a small pressure of 0.5\,GPa, while both SC phases remain unaltered, the same SC order parameter should exist in the tetragonal as well as in the orthorhombic phase.

\begin{acknowledgments}
We thank Thomas Lorenz for having inspired this work. We are indebted to P. Brydon, J. Landaeta, S. Galeski, C. Lee, R. Ohlendorf, B. Schmidt, J. Sourd, Y. Song, Y. Liu, C. Cao, and M. Grosche for inspiring discussions. AL thanks financial support from ANID-FONDECYT Initiation grant N. 11251906. We acknowledge support from Deutsche Forschungsgemeinschaft (DFG) through W\"{u}rzburg-Dresden Cluster of Excellence on Complexity, Topology and Dynamics in Quantum Matter — ctd.qmat (EXC 2147, project ID 390858490) and from HLD at HZDR, member of the European Magnetic Field Laboratory (EMFL). 
\end{acknowledgments}

\appendix
%

%
\end{document}


\preprint{APS/123-QED}
\title{Supplemental Material \\ Tetragonal-to-orthorhombic distortion in CeRh$_{2}$As$_{2}$}

\author{P. Khanenko}
\email[Corresponding author:~]{p.khanenko@hzdr.de}
\affiliation{Hochfeld-Magnetlabor Dresden (HLD-EMFL) and Würzburg-Dresden Cluster of Excellence ctd.qmat, Helmholtz-Zentrum Dresden-Rossendorf, 01328 Dresden, Germany}
\affiliation{Max Planck Institute for Chemical Physics of Solids, 01187 Dresden, Germany}
\author{M. Pfeiffer}
\affiliation{Max Planck Institute for Chemical Physics of Solids, 01187 Dresden, Germany}
\affiliation{Institute for Solid State and Materials Physics, Technical University Dresden, 01062 Dresden, Germany}
\author{K. Semeniuk}
\affiliation{Max Planck Institute for Chemical Physics of Solids, 01187 Dresden, Germany}
\affiliation{Institute for Quantum Materials and Technologies, Karlsruhe Institute of Technology, 76131 Karlsruhe, Germany}
\author{E. Hassinger}
\affiliation{Max Planck Institute for Chemical Physics of Solids, 01187 Dresden, Germany}
\affiliation{Institute for Quantum Materials and Technologies, Karlsruhe Institute of Technology, 76131 Karlsruhe, Germany}
\author{P. Thalmeier}
\affiliation{Max Planck Institute for Chemical Physics of Solids, 01187 Dresden, Germany}
\author{J. Wosnitza}
\affiliation{Hochfeld-Magnetlabor Dresden (HLD-EMFL) and Würzburg-Dresden Cluster of Excellence ctd.qmat, Helmholtz-Zentrum Dresden-Rossendorf, 01328 Dresden, Germany}
\affiliation{Institute for Solid State and Materials Physics, Technical University Dresden, 01062 Dresden, Germany}
\author{A. M. Le\'on}
\affiliation{Departamento de Física, Facultad de Ciencias, Universidad de Chile, Casilla 653, Santiago, Chile}
\author{G. Zwicknagl}
\affiliation{Institute for Mathematical Physics, Technische Universit\"at Braunschweig, D-38106 Braunschweig, Germany}
\affiliation{Max Planck Institute for Chemical Physics of Solids, 01187 Dresden, Germany}
\author{C. Geibel}
\affiliation{Max Planck Institute for Chemical Physics of Solids, 01187 Dresden, Germany}
\author{S. Khim}
\affiliation{Max Planck Institute for Chemical Physics of Solids, 01187 Dresden, Germany}
\author{M. Brando}
\email[Corresponding author:~]{brando@cpfs.mpg.de}
\affiliation{Max Planck Institute for Chemical Physics of Solids, 01187 Dresden, Germany}
\date{\today}

\maketitle
\section{Influence of magnetic field on thermal expansion in phase II}
\begin{figure}[b]
	\begin{center}
		\includegraphics[width=\columnwidth]{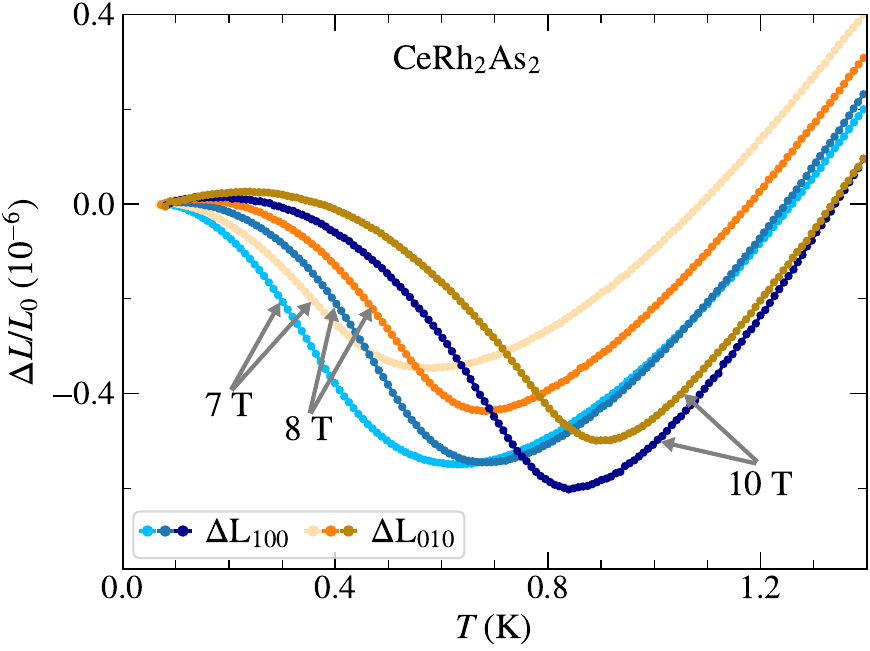}
		\caption{Longitudinal thermal expansion $\Delta L(T)/L_{0}$ along the crystallographic directions [100] and [010] at 7, 8, and 10\,T.}
		\label{figS2}
	\end{center}
\end{figure}

Figure~\ref{figS2} shows the longitudinal thermal expansion along the directions [100] and [010] at 7, 8, and 10\,T. At 7\,T, a clear difference is observed between $\Delta L_{[100]}$ and $\Delta L_{[010]}$,  even though no such difference is expected in a tetragonal system. This difference can be attributed to domains that exist within phase II, the nature of which is yet to be determined. The difference between $\Delta L_{[100]}$ and $\Delta L_{[010]}$ decreases with increasing field, as is shown in Fig.~\ref{figS2}, suggesting that the magnetic field supports the formation of a single-domain state within phase II. However, the difference may not disappear completely due to a misalignment of the sample.
\section{Ehrenfest relation for phase II}
\begin{figure*}[t]
	\begin{center}
		\includegraphics[width=\textwidth]{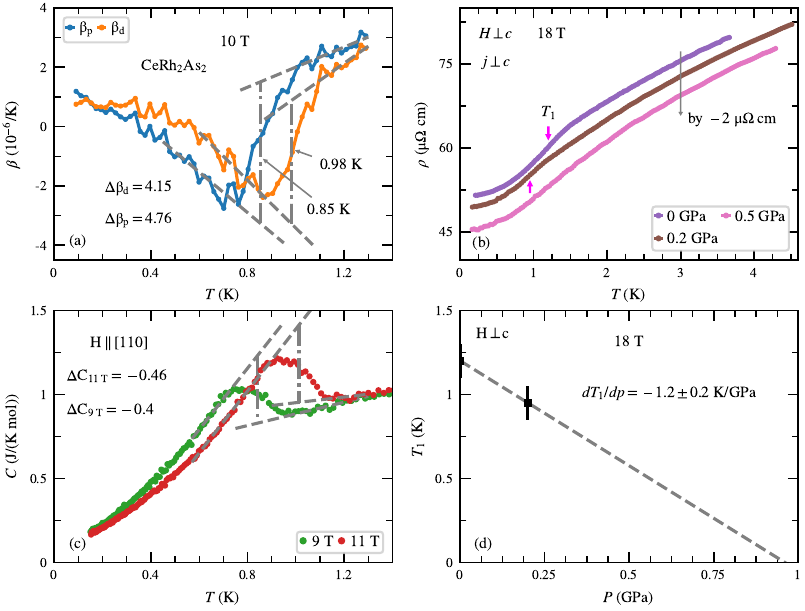}
		\caption{(a) Volumetric thermal-expansion coefficient $\beta(T)$, calculated using data shown in Fig.\,4(b) of the main paper. $\beta_{\mathrm{p}} = \alpha_{001} + \alpha_{100} + \alpha_{010}$, $\beta_{\mathrm{d}} = \alpha_{001} + \alpha_{110} + \alpha_{1\overline{1}0}$, $\alpha_{x}$ is the longitudinal thermal-expansion coefficient along a direction $x$. The thermal-expansion coefficient along [001] is taken from Ref.~\cite{khanenko2025t}. (b) Electrical resistivity $\rho$ versus temperature at various hydrostatic pressures. The curves are shifted for clarity. The samples were not oriented in the $ab$ plane. Magenta arrows indicate the PM-to-phase II transition. (c) Specific heat versus temperature of a sample of comparable quality. (d) Temperature vs. pressure phase diagram of CeRh$_{2}$As$_{2}$ for a magnetic field of 18\,T applied perpendicular to the $c$ axis, constructed using the data shown in panel (b).}
		\label{figS1}
	\end{center}
\end{figure*}

To calculate the expected pressure dependence of the transition temperature $T_{\textrm{1}}$ to phase II, we use the Ehrenfest relation for a second-order phase transition,  $dT_{1}/dp = V_{m}T_{1}(\Delta \beta/\Delta C)$, where $V_{m}$ is the molar volume, $\Delta \beta$ is the jump in the volumetric thermal-expansion coefficient and $\Delta C$ is the jump in the specific heat at $T_{\textrm{1}}$. Due to the in-plane anisotropy of the $T - H_{ab}$ phase diagram, two volumetric thermal-expansion coefficients were calculated, $\beta_{\mathrm{p}} = \alpha_{001} + \alpha_{100} + \alpha_{010}$, $\beta_{\mathrm{d}} = \alpha_{001} + \alpha_{110} + \alpha_{1\overline{1}0}$. Here, $\alpha_{x}$ is the longitudinal thermal-expansion coefficient along the $x$ direction. $\beta_{\mathrm{p}}$ and $\beta_{\mathrm{d}}$ are shown in Fig.~\ref{figS1}(a). We used the transverse thermal expansion along the $c$ axis, with the field applied along [110] reported in Ref.~\cite{khanenko2025t} to calculate $\alpha_{001}$. The specific heat $C$ of a CeRh$_{2}$As$_{2}$ sample of comparable quality is shown in Fig.~\ref{figS1}(c). As no measurements were conducted at 10\,T, we used the jump in $C$ at 9 and 11\,T at the transition to phase II to calculate the expected jump at 10\,T, $\Delta C _{\mathrm{10\ T}} = -0.43$\,$\mathrm{J/(K \ mol})$. Based on previous measurements~\cite{khanenko2025a}, we expect a linear change in $\Delta C$ in magnetic field. Using these values, we obtain $dT_{1}/dp = V_{m}T_{1p}(\Delta \beta_{\mathrm{p}}/\Delta C_{\mathrm 10\ T}) = (-0.5 \pm 0.1$\,$\mathrm{K})/\mathrm{GPa}$ and $dT_{1}/dp = V_{m}T_{1d}(\Delta \beta_{\mathrm{d}}/\Delta C_{\mathrm 10\ T}) = (-0.5 \pm 0.1$\,$\mathrm{K})/\mathrm{GPa}$, where $T_{1p}$ and $T_{1d}$ are $T_{1}$ for the directions [100] and [110], respectively. Surprisingly, the values of $dT_{1}/dp$ are the same for both field directions.

To verify the expected pressure dependence of $T_{\textrm{1}}$, we conducted electrical-resistivity measurements under hydrostatic pressure and an in-plane magnetic field of 18\,T, shown in Fig.~\ref{figS1}(b). Further details of the experiment can be found in the SM of Ref.~\cite{pfeiffer2024}. As the samples were not oriented in the $ab$ plane, it is unclear along which in-plane direction the current and magnetic field were applied. We determined the transition temperature $T_{\textrm{1}}$ from the derivative of the resistivity. At 0.5\,GPa, the signature of $T_{\textrm{1}}$ broadens significantly, and the transition temperature cannot be determined unambiguously. From the extracted transition temperatures, we constructed the temperature vs. pressure phase diagram, shown in Fig.~\ref{figS1}(d). We found that $T_{\textrm{1}}$ is suppressed by hydrostatic pressure at a rate more than twice that calculated,  $dT_{1}/dp = -1.25 \pm 0.2$\,$\mathrm{K/GPa}$. While such discrepancy in the absolute values is not uncommon, there is a qualitative agreement between the experiment and the calculation. Therefore, it is unclear whether the tetragonal symmetry is preserved at the PM-to-phase II transition.
\section{Theory Calculations}
\subsection{Computational Methods}
Electronic-structure calculations were performed within density-functional  theory (DFT) using the Vienna \textit{ab initio} simulation package (VASP)  and the PBEsol exchange-correlation functional. A plane-wave cutoff of 
500~eV and Monkhorst--Pack $k$-point meshes of $12\times12\times6$ and $16\times16\times6$ were used for structural relaxations and self-consistent calculations, respectively. Starting from the experimental lattice parameters, the internal atomic positions were relaxed until the forces were below $10^{-3}$~eV/\AA.

To investigate the interplay between magnetism and structural degrees of freedom, different magnetic-moment orientations, $[100]$, $[110]$, and $[001]$, were considered including spin--orbit coupling. Electronic correlations were treated within the Dudarev DFT+$U$ approach, with $U_{\mathrm{eff}}=0.5$, $1$, $1.5$, and $2$~eV, allowing us to examine the 
effect of localization and magnetic-moment magnitude.

\subsection{Results}
\begin{figure}[t]
    \centering  
    \hspace*{-0.7cm}
    \includegraphics[width=\columnwidth]{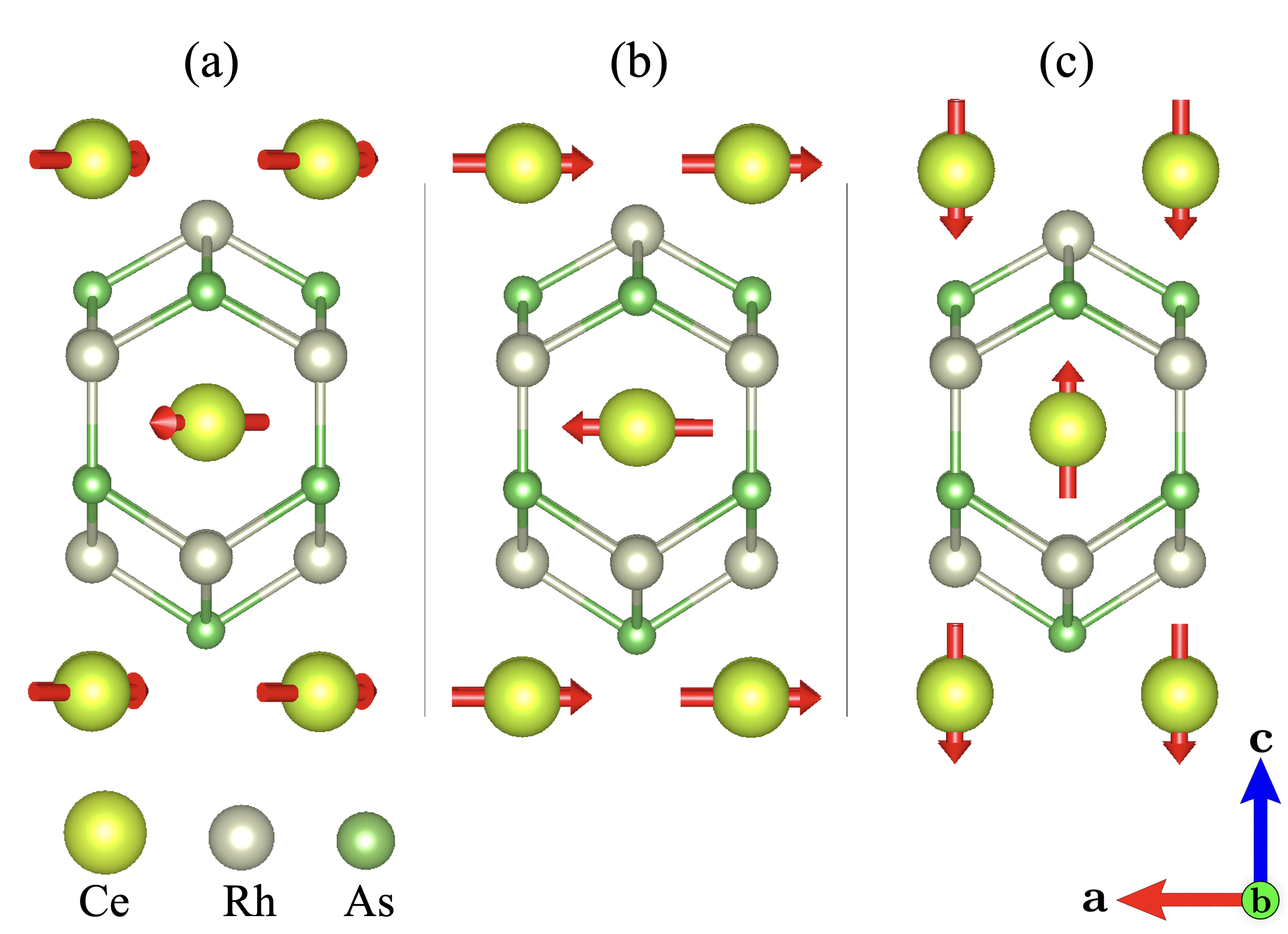}
    \caption{Magnetic configurations considered in the DFT+SOC calculations, with the Ce magnetic moments aligned along (a) [110], (b) [100], and (c) [001].}
    \label{figS3}
\end{figure}

To investigate the interplay between the structural distortions observed in CeRh$_2$As$_2$ and the magnetic configuration, we performed DFT+$U$  calculations including spin--orbit coupling (SOC), considering the Ce magnetic 
moments oriented along [100], [110], and [001], as shown in Fig.~\ref{figS3}.
\begin{figure}[b]
    \centering
    \includegraphics[width=\columnwidth]{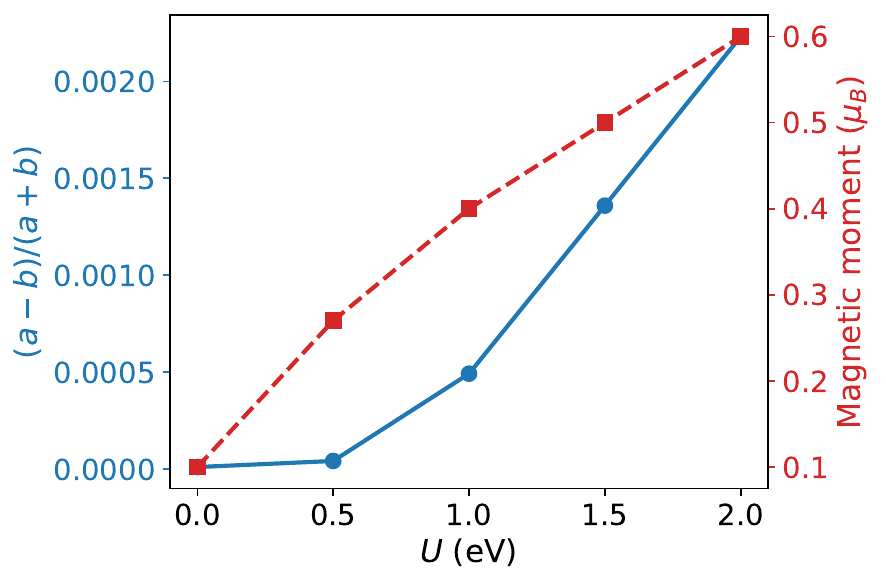}
    \caption{Orthorhombic distortion, $\delta=(a-b)/(b+a)$, as a function of the Hubbard-$U$ parameter for calculations with the Ce magnetic moments along the [100] direction.}
    \label{figS4}
\end{figure}

Our results reveal a clear dependence of the relative stability of the different magnetic configurations on the Hubbard-$U$ parameter. At $U=0.5$~eV, the configuration with the Ce magnetic moments oriented along [110] is energetically favored over the [100] configuration by approximately 0.2~meV. At $U=1$~eV, this energy hierarchy is reversed, with the [100] orientation becoming more stable by approximately 5~meV, and this trend is maintained up to $U=2$~eV. This change in relative stability occurs concomitantly with a spontaneous structural change from the tetragonal to an orthorhombic structure characterized by a $B_{1g}$-type distortion, as shown in Fig.~\ref{figS4}. In contrast, the configuration with the magnetic moments oriented along [001] remains the highest in energy over the explored range of Hubbard-$U$ values.
\subsection{Renormalized Band Calculations}
We consider the sensitivity of the renormalized bands in CeRh$_{2}$As$_{2}$ to a perturbation with $B_{2g}$ symmetry at the Ce sites. Its action on the Ce 4$f$ states is accounted for by the Hamiltonian
\begin{equation}
    H_{xy} = \sum_{\mathbf{R}_i} \sum_{\alpha=1,2} \sum_{\Gamma \Gamma'} \lambda_{xy}^{\alpha} f_\Gamma^\dagger(\mathbf{R}_{i\alpha}) \left(O_{xy}\right)_{\Gamma\Gamma'} f_{\Gamma'}(\mathbf{R}_{i\alpha}) \, .
    \label{eq1}
\end{equation}
As written, $f_\Gamma^\dagger(\mathbf{R}_{i\alpha})$ and $f_{\Gamma'}(\mathbf{R}_{i\alpha})$ are the creation and annihilation operators for a 4$f$ electron in CEF states $\Gamma$, $\Gamma'$ at the Ce atom in the unit cell $i$, with label $\alpha =1, 2$, and $(O_{xy})_{\Gamma\Gamma'}$ are the matrix elements of the operator $\hat{O}_{xy}$ in the basis of the CEF states. The indices $\Gamma$ and $\Gamma'$ are a short-hand notation for a CEF state $\vert \Gamma \rangle =\vert \Gamma_7^{(1,2)}\tau \rangle \,\,\vert \Gamma_{6}\tau \rangle$ of symmetry $\Gamma_{7}$, $\Gamma_{6}$ and Kramers-degeneracy index $\tau = \pm$. The overall strength of the homogeneous perturbation is given by the parameter $\lambda^{\alpha}_{xy}$. Of particular interest are the uniform (“ferro”) case with $\lambda_{1xy} = \lambda_{2xy}$ and the staggered (“anti-ferro”) case with $\lambda_{1xy} = -\lambda_{2xy}$. After Fourier transformation with respect to the lattice vectors $\mathbf{R}_{i}$ we obtain

\begin{equation}
    H_{xy} = \sum_{\mathbf{k}} \sum_{\alpha=1,2} \sum_{\Gamma \Gamma'}\lambda_{xy}^{\alpha} f_{\alpha\mathbf{k}\Gamma}^\dagger \left(O_{xy}\right)_{\Gamma\Gamma'} f_{\alpha\mathbf{k}\Gamma'} \, .
    \label{eq2}
\end{equation}

The periodic perturbation shifts the energies of the quasiparticle states $|n \textbf{k}\rangle$ with band index $n$ and wave vector $\textbf{k}$ and induces inter-band transitions. To first order in $\lambda^{\alpha}_{xy}$, the energy change is given by the $\textbf{k}$-dependent expectation value

\begin{equation}
    \delta E \left(n\mathbf{k}\right) = \langle n\mathbf{k} | H_{xy} | n\mathbf{k} \rangle \, .
    \label{eq:dE}
\end{equation}

For the uniform distortion $\lambda_{xy}^1 = \lambda_{xy}^2 = \lambda_{xy}^u$, the energy shift is given by the expectation value 

\begin{equation}
    \delta E^u \left(n\mathbf{k}\right)= \lambda_{xy}^u \left\langle n\mathbf{k} \left| \sum_{\alpha} \sum_{\Gamma \Gamma'} f_{\alpha\mathbf{k}\Gamma}^\dagger \left(O_{xy}\right)_{\Gamma\Gamma'} f_{\alpha\mathbf{k}\Gamma'} \right| n\mathbf{k} \right\rangle \, .
    \label{eq:dEu}
\end{equation}

The influence of a perturbation with $B_{1g}$ symmetry is obtained in close analogy by replacing 

\begin{equation}
     \lambda_{xy} \left(O_{xy}\right)_{\Gamma\Gamma'} \rightarrow  
     \lambda_{x^2-y^2} 
     \left(O_{x^2-y^2 }\right)_{\Gamma\Gamma'}
     \, .
    \label{eq:B1g}
\end{equation}

Fig.\,5 in the main text displays the sensitivity $\left\langle n\mathbf{k} \left| \sum_{\alpha} \sum_{\Gamma \Gamma'} f_{\alpha\mathbf{k}\Gamma}^\dagger \left(O_{xy}\right)_{\Gamma\Gamma'} f_{\alpha\mathbf{k}\Gamma'} \right| n\mathbf{k} \right\rangle \ $ for the states at the
Fermi energy in the basal and in the top plane with respect to a uniform (“ferro”) perturbation with $B_{2g}$ (top panels) and $B_{1g}$ (bottom panels) symmetries.

The energy shift $\delta E^u $ leads to a distortion of the constant-energy surfaces. The Fermi wave vector of the undistorted lattice will change, $\mathbf{k}_F(n) \rightarrow \mathbf{k}_F(n) + \delta \mathbf{k}_F(n)$, where the shift $ \delta \mathbf{k}_F(n)$ satisfies

\begin{equation}
    \delta E^u(n\mathbf{k}) \simeq \hbar \mathbf{v}_F(n) \cdot \delta \mathbf{k}_F(n),
    \label{eq:dEu_kF}
\end{equation}

with the Fermi velocity $\mathbf{v}_F(n)$.

\appendix
%

%